# Trapping neutral atoms in the field of a vortex pinned by a superconducting nano-disc

Vladimir Sokolovsky, Daniel Rohrlich, Baruch Horovitz

*Department of Physics, Ben Gurion University of the Negev, Beersheba 84105, Israel*

**Abstract**

Atom chips made of superconducting material can generate magnetic traps with significantly reduced noise. Recently, several designs for superconducting chips have been theoretically analyzed and experimentally tested, for cases with many vortices considered as an average vortex density. Here we show theoretically, for the first time, how the magnetic field of a *single* vortex, pinned by a superconducting nano-disc of radius ~100 nm and combined with an external bias field parallel to the disc surface, yields a closed 3D trap for cold atoms. The size of the trap, and its height above the superconductor surface, are typically tens or hundreds of nanometers. We estimate the average lifetime $\tau$ of $^{87}$Rb (rubidium) atoms (subject to thermal escape and Majorana spin flips) in the range 0.05-1.0 ms. Next, we model the trap in a quantum adiabatic approximation and apply Fermi's rule to estimate the lifetime of $^{87}$Rb atoms in the ground state of this trap. We obtain similar lifetimes $\tau$ as in the semiclassical estimate, in the range 0.05-3.5 ms. We find that $\tau$ depends on the gradient $B_0$ of the vortex's magnetic field according to $\tau \sim B_0^{-2/3}$.

## 1. Introduction

The application of superconductors to atom chips is a recent development that presents new opportunities for atom optics. One advantage of superconductors over conventional conductors is a significant enhancement of trapping lifetime: atoms escape magnetic traps when their spins flip, and the reduced noise of superconducting atom chips leads to a reduction in spin flips. A major experimental goal is to trap cold atoms within a micron of the chip surface. Magnetic fluctuations near a metallic surface induce spin-flip transitions to untrapped magnetic sublevels and thus to significant loss of atoms from the trap [1]. Theoretical studies of superconducting atom chips [2-4] predict an impressive reduction in noise of 6-12 orders of magnitude. The reduction is predicted to be most significant when the atom's distance $z$ from the chip surface is in the range $\lambda < z < \delta_{skin}$, where $\lambda$ is the London penetration length and $\delta_{skin}$ is the skin depth of the normal phase; e.g. for Rb atoms above an Nb chip, with a spin-flip energy corresponding to 560 kHz, we have $\lambda = 35$ nm and $\delta_{skin} = 150$ $\mu$m. Thus Ref. [3] predicts a lifetime of 5000 s at a trap height of 1 $\mu$m; by contrast, in a trap of the same height above a normal metal, at room temperature, the lifetime is less than 0.1 s [5]. Yet experimental data [6-7] from superconducting chips with $z = 30$ $\mu$m show an enhancement of the lifetime of only one order of magnitude, indicating that additional sources of noise reduce the lifetime. An additional source of magnetic noise may be the fluctuations of isolated vortices [8].

The properties of magnetic atom traps over superconducting chips have been theoretically investigated in Refs. [9-13]. In these papers, the specific properties of superconductors in both the Meissner state and the mixed state (where magnetic flux partially penetrates the superconductor in the form of a vortex lattice) were considered. These traps also decrease technical noise via use of a persistent current or trapped magnetic flux [6, 11-15]. Realizations of atom chips with superconducting elements have been reported in Refs. [1,6,14-22]. The first experiment, by Nirrengarten et al. [16], demonstrated the advantages of superconducting chips over normal-metal chips. Current passed through niobium wires (in both the "U" and "Z" trap configurations) cooled to about 4.2 K. The resulting atom spin relaxation time (lifetime) was estimated at 115 s. This value is comparable to the best achieved for atoms trapped near normal-metal wires [23]. A further result was achieved in Ref. [1], where the authors reported an estimated lifetime of 10 minutes in a magnetic trap 300 μm above an atom chip consisting of a niobium strip covered by a gold layer. These experiments [1,6,14-22] showed the possibility both of creating superconducting magnetic traps for cold atoms and of investigating their superconducting properties via the atom traps, including the stability of magnetic hysteresis [15], memory effects [20], temperature of dendritic instability [19], and the influence of laser radiation on the critical current [17].

All the superconducting chips experimentally tested and/or theoretically considered in these works operate in the mesoscopic limit, in which characteristic lengths are comparable to or larger than the average vortex separation. Thus the calculated magnetic fields refer to average vortex density rather than to individual vortices, except in Ref. [13] where distances become comparable to the vortex spacing. (The typical sizes of superconducting wires and of the atom cloud are between several tens to hundreds of μm; a vortex diameter is determined by the London penetration depth $\lambda$ which is of the order of 100 nm). Here we present the first theoretical prediction of trapping of cold atoms by nano-scale magnetic traps obtained by combining the magnetic field of a vortex with an external DC bias field. In contrast to Refs. [1,6,9-22], we consider the magnetic field due to the currents of a *single* vortex rather than of a vortex lattice.

This paper is organized as follows. Section 2 presents a detailed solution of the single-vortex trap, from the structure of the vortex and its magnetic field to the addition of a constant bias field and the analysis of the trap thereby created, with an estimate of the trap depth. Section 3 obtains energy levels for neutral atoms in the trap, and calculates their lifetimes as a function of their temperature. The characteristic lifetime is 0.05-1.0 ms. However, the treatment in this section is semiclassical. The treatment in Sect. 4 is quantum and leads to an effective adiabatic Hamiltonian for the atoms. Applying Fermi's golden rule, we estimate the lifetime of the atoms in the trap at zero temperature up to 3.5 ms. Thus, the close proximity of the atoms to the chip (tens or hundreds of nm) comes at the price of short trapping time. We end with a brief Conclusion.

## 2. Magnetic trap of a single vortex

Let us consider a disc-shaped type-II superconducting film, of radius $R$ and thickness $\delta \ll R, \lambda$ in the *x-y* plane, containing one vortex at its center (at the origin $x = y = z = 0$). In type-II superconductors, for Ginzburg-Landau parameter $k = \lambda / \xi \gg 1$ (where $\xi$ is the coherence length), the core of a vortex of radius close to $\xi$ can be neglected, and the magnetic vector potential **A** of a single straight vortex along the *z*-axis should satisfy the modified London equations [24-25]:

$$\nabla \times \nabla \times \mathbf{A} = \mu_0 \mathbf{j} \quad , \tag{1}$$

$$\mathbf{j} = \frac{1}{\mu_0 \lambda^2}\left[\frac{\Phi_0}{2\pi r}\hat{\boldsymbol{\varphi}} - \mathbf{A}(r,z)\right] ,\qquad(2)$$

where $\hat{\boldsymbol{\varphi}}$ is the azimuthal unit vector and $\nabla \times \nabla \times \mathbf{A} = 0$ in the vacuum. Here $\mu_0$ is the magnetic permeability of vacuum and $\Phi_0 = 2.07 \times 10^{-15}$ G·cm² is the quantum of magnetic flux. The vector potential satisfies the Coulomb gauge condition, $\nabla \cdot \mathbf{A} = 0$, and vanishes on the vortex axis [26].

The current density $\mathbf{j}$ and vector potential $\mathbf{A}$ possess only angular components in the cylindrical coordinate system $r$, $\varphi$, $z$; we denote them by $j$ and $A$, respectively. The magnetic field of the disc will be determined as the superposition of magnetic fields created by the current elements $di = j\,dz\,dr$ in rings of radius $r_\circ \leq R$, height $dz$, and thickness $dr$. The vector potential $A_{ring}$ and the $r$- and $z$-components $B_{ring,r}$ and $B_{ring,z}$ of the magnetic field of a ring with negligible cross-section of the wire are [27]

$$dA_{ring}(r,z,r_\circ) = G_A\,di \equiv \frac{\mu_0 di}{\pi\sqrt{m}}\sqrt{\frac{r_\circ}{r}}\left[(1-0.5m)K(m) - N(m)\right] ,\qquad(3)$$

$$dB_{ring,r}(r,z,r_\circ) = G_{B,r}\,di \equiv \frac{\mu_0 di}{2\pi r}\frac{z}{\sqrt{(r+r_\circ)^2+z^2}}\left[-K(m) + \frac{r^2+r_\circ^2+z^2}{(r_\circ-r)^2+z^2}N(m)\right] ,\qquad(4)$$

$$dB_{ring,z}(r,z,r_\circ) = G_{B,z}\,di \equiv \frac{\mu_0 di}{2\pi}\frac{1}{\sqrt{(r+r_\circ)^2+z^2}}\left[K(m) + \frac{r^2-r_\circ^2-z^2}{(r_\circ-r)^2+z^2}N(m)\right] ,\qquad(5)$$

where $m = \dfrac{4rr_\circ}{(r+r_\circ)^2+z^2}$ and $K(m) \equiv \int_0^{\pi/2}\dfrac{d\beta}{\sqrt{1-m\sin^2\beta}}$, $N(m) \equiv \int_0^{\pi/2}\sqrt{1-m\sin^2\beta}\,d\beta$ are the complete elliptic integrals of the first and second kinds, respectively. The vector potential of the disc can be written as

$$A(r,z) = \int_0^R dr_\circ \int_{-\delta/2}^{\delta/2} j(z',r_\circ)G_A(r,z-z',r_\circ)\,dz' .\qquad(6)$$

For a thin disc, $(\delta/R)^2 \ll 1$, the vector potential in the superconductor, $-\delta/2 \leq z \leq \delta/2$, can be presented as a function of $r$ only:

$$A = \int_0^R J(r_\circ)G_A(r,0,r_\circ)\,dr_\circ ,\qquad(7)$$

where $J = \int_{-\delta/2}^{\delta/2} j\,dz$ is the sheet current density. Integrating (2) over $z$ from $-\delta/2$ to $\delta/2$ and using the normalized dimensions, we obtain the following integral equation:

$$\tilde{J}(\rho) = -\frac{R\delta}{\pi\lambda^2} \int_0^1 \frac{\tilde{J}(\rho_\circ)}{\sqrt{m_1}} \sqrt{\frac{\rho_\circ}{\rho}} \left[(1-0.5m_1)K(m_1) - N(m_1)\right] d\rho_\circ + \frac{1}{\rho} \quad , \tag{8}$$

where $\rho = r/R$, $\rho_\circ = r_\circ/R$, $m_1 = \dfrac{4\rho\rho_\circ}{(\rho+\rho_\circ)^2}$, and $\tilde{J}$ is the sheet current density normalized (divided) by $\dfrac{\delta}{\mu_0\lambda^2}\dfrac{\Phi_0}{2\pi R}$. We consider in particular the radius $R \approx \lambda$.

Equation (8) contains a single parameter $\varepsilon = R\delta/\pi\lambda^2$, and its solution can be sought numerically or via the series $\tilde{J}(\rho) = \sum_{n=0} \varepsilon^n \tilde{J}_n(\rho)$. From Eq. (8) we have

$$\tilde{J}_0(\rho) = \frac{1}{\rho} \quad , \tag{9}$$

and

$$\tilde{J}_n(\rho) = -\int_0^1 \frac{\tilde{J}_{n-1}(\rho_\circ)}{\sqrt{m_1}} \sqrt{\frac{\rho_\circ}{\rho}} \left[(1-0.5m_1)K(m_1) - N(m_1)\right] d\rho_\circ \tag{10}$$

for $n > 0$. In the case of a thin disc, $\varepsilon \ll 1$, it is enough to take into account only the first several terms in the series. Let us seek $\tilde{J}_1$. Note that at $\rho_\circ \ll 1$, we have

$$\frac{1}{\rho_\circ \sqrt{m_1}} \sqrt{\frac{\rho_\circ}{\rho}} \left[(1-0.5m_1)K(m_1) - N(m_1)\right] \approx \frac{\pi\rho_\circ}{4\rho^2} \quad . \tag{11}$$

Hence, the only singularity in (10) is the integrable singularity of $K(m_1)$ at $m_1 \to 1$, i.e. at $\rho_\circ \to \rho$. In the limit $m_1 \to 1$, we can apply [28] the result $K(m_1) = \dfrac{1}{2}\log\left(\dfrac{16}{1-m_1}\right)$; changing variables and passing from integration over $\rho_\circ$ to integration over $t^2 = 1 - m_1$ leaves us with an expression containing a singularity of the type $\log(t)$ when one of the integration limits equals zero. A good fit to the numerical integration is $\tilde{J}_1(\rho) \approx -1.565 + 0.8\rho + 0.193\rho^6$. The error in this fit does not exceed 1% except in the region near zero, $\rho < 0.005$, where the result of the integration increases to zero but the fit gives about $-1.565$. However, contributions of currents in this area to the vector potential and magnetic field are negligibly small. The solution of Eq. (10) to order $\varepsilon^2$ is

$$\tilde{J}(\rho) = \frac{1}{\rho} - \varepsilon\left(1.565 - 0.8\rho - 0.193\rho^6\right) + \ldots \quad . \tag{12}$$

In the superconductor, the dimensionless potential vector is $\tilde{A} = \dfrac{2\pi R}{\Phi_0} A = -\varepsilon \tilde{J}_1$. To a first approximation, the solution obtained implies that the current density is proportional to $1/r$, as obtained in [24] for a thin superconducting disc, in [29] for a thin infinite film at $r \ll 2\lambda^2/\delta$, and in [26,30] for a bulk superconductor at $r \ll \lambda$. In [24], to find the vector potential, Eqs. (1-2) were reduced to an integral equation for a vector potential, and it was shown that $\tilde{A} \sim \dfrac{R\delta}{2\lambda^2}$; in our case $\tilde{A} \sim \dfrac{R\delta}{\pi\lambda^2}$, i.e. in the two cases the small parameters are about the same: they differ by a coefficient of about 1.5.

The magnetic field of a trap obtained from a single vortex pinned to a superconducting disc, combined with a bias magnetic field parallel to the *x*-axis, is presented in Fig. 1. We plot the magnitude $B_{tot}$ of the total field (a vector sum of the bias $\mathbf{B}_{bias}$ and vortex fields), since it is this total field magnitude that creates magnetic trapping, in the adiabatic approximation. There are minima in $B_{tot}$ in three planes; that is, the total field yields a closed 3D magnetic trap. The magnetic trap was calculated using the first approximation for the sheet current density, Eq. (9), and the vortex field is determined by integration of Eqs. (4-5) over $r_\circ$. The results are presented in a dimensionless form: the magnetic field is normalized by $B_{norm} = \Phi_0 \delta / 2\pi\lambda^2 R$, and we define $\tilde{x} = x/R$, $\tilde{y} = y/R$, $\tilde{z} = z/R$.

The coordinates of the trap center as a function of the bias field are presented in Fig. 2. An increase in the bias field leads, as in the case of the "side-guide" configuration [10], to a decreased trap height (the *z*-coordinate of the trap center). At the same time, the increase moves the trap center towards the disc axis: at low bias fields, the trap center is not above the superconductor and moves above it only for $\tilde{B}_{bias} > 0.13$. At any $\tilde{z}$ the *z*-component of the vortex magnetic field decreases with an increase in radius and changes sign at some point $\rho(\tilde{z})$, while $B_r$ does not change its sign at $\tilde{z} > 0$, where the trap is analyzed. The coordinates of the trap center are determined from the conditions $B_z = 0$ and $(B_x, B_y, 0) + \mathbf{B}_{bias} = 0$.

The trap center is the position at which $\mathbf{B}_{tot} = \mathbf{0}$, i.e. $\mathbf{B}_{tot}$ changes sign along any path traversing the trap center. If the spin of a moving atom follows the direction of $\mathbf{B}_{tot}$, it is trapped by the magnetic field. Nonadiabatic effects can, however, induce (Majorana) spin flips. We consider such effects in Sects. 3-4. If $\mathbf{B}_{bias}$ is directed along the *x*-axis, the minimum will be at $\tilde{y} = 0$ where the *y*-component of the total field is zero. To reduce spin-flip losses of the trapped atoms, an additional magnetic field perpendicular to the bias field is usually applied [10,16,31]. In our case, application of additional magnetic fields along the *y*- or *z*-axis moves the trap center but does not increase the field magnitude at the trap center, i.e. $\mathbf{B}_{tot} = 0$ at the center.

Two figures of merit are commonly used for describing the confinement of cold atoms in a magnetic trap: the magnetic gradient at the trap center, and the depth of the trapping potential. The trap depth is determined as the (total) potential barrier at its minimal height, from the trap center either to the superconductor surface or away from it. Our calculation shows that the minimal height of the potential barrier in the *x*- and *y*-directions is achieved away from the trap center, and equals the bias field. In the *z*-direction, at low bias fields $\tilde{B}_{bias} \leq 0.16$, the minimal height is also achieved away from the trap center and equals the bias field (Fig. 3); at higher fields, the minimal height is achieved at the superconductor surface. In the calculation, the surface is at $\tilde{z} = 0.15$. The dependence of the trap depth and height on the bias field is similar to their dependence in the side-guide configuration [10]: as the bias field increases, the depth increases to the maximum and then decreases, while the trap height decreases monotonically. Near the surface, the trap depth is insufficient for stable trapping. For the side-guide chip, this result was theoretically predicted [10] and experimentally confirmed [20].

To analyze the possibility of an atom trap based on a single vortex, we use reliability criteria for trapping of atoms at a representative temperature of 1 μK: the trap depth should exceed 10 μK (i.e. 10 times the temperature of the atoms) and the gradient should be high enough to overcome the acceleration of gravity. Below we analyze trap stability for the atoms in the state with $F=2$, $m_F=1$ where $F$ is the total spin and $m_F$ is its projection onto the local magnetic field. The trap depth and gradient are 0.16 G and 30 G/cm, respectively. Let us estimate the value of $B_{norm}$ used to normalize the magnetic field. The London penetration depth for the type-II superconductors depends on many factors: type of superconductor, preparation technology, temperature, etc. For example, the depth for $Nb_3Sn$ is 65 nm; for $MgB_2$ film, about 110 nm [33]; and for YBCO film, about 200 nm [34] at zero temperature. For $R = \lambda = 100$ nm and $\delta = 0.3R$, the value of $B_{norm}$ is estimated as 100 G; then $\tilde{B}_{bias} = 0.0016$ is the minimal normalized bias field yielding a trap with the required potential barrier. Calculations show that the minimal gradient in all directions is about the same and increases with the bias field, $\sim B_{bias}^{1.25}$. For example, at $\tilde{B}_{bias} = 0.006$ (see Fig. 4), the average gradient near the trap center can be estimated as $2 \times 10^4$ G/cm, which satisfies the above-mentioned reliability criteria [10,12,31] by three orders of magnitude. The characteristic trap size should be about $2R = 0.2$ μm and decrease with an increase of the bias field. Modern technology allows production of superconducting thin film structures with characteristic size ~25 nm [35]. Planar superconducting structures containing the discs, each of which pins a vortex, can be designed for creation of a set of the nano-traps. The distance between the neighboring traps can be decreased to a few hundred nanometers. In the approximation of an infinitely thin superconductor, only the $z$-component of an external magnetic field influences the current distribution. Outside the disc, this component of the vortex field decreases rapidly with an increase of the distance $\rho$ from disc axis, and at $\rho > 2$ the field of a neighboring disc can be neglected in comparison with the self-field. The current distribution in each disc can be calculated separately.

## 3. Trap stability: semiclassical treatment

In this section we consider the thermodynamics of trapped atoms, in the semiclassical adiabatic approximation that atom magnetic moments (spins) always line up with the magnetic field so as to minimize the energy. We also derive semiclassical criteria for the adiabatic approximation. The next section considers the quantum criteria for the adiabatic approximation.

The motion of an atom in a magnetic trap can be described by the Schrödinger equation for the wave function $\psi$:

$$E\psi = -\frac{\hbar^2}{2m}\nabla^2\psi + V\psi \quad , \qquad (13)$$

where $m$ and $E$ are the mass of an atom and its energy, **S** (with half-integer eigenvalues) is the spin vector, $V = \mu_B \mathbf{S} \cdot \mathbf{B}_{tot}$ is the potential energy of an atom in the magnetic field $\mathbf{B}_{tot}$ of the trap, and $\mu_B \mathbf{S}$ is the atom magnetic moment. The numerical calculation of Sect. 2 shows that the magnetic field grows linearly near the trap center:

$$\mathbf{B}_{tot} = (a_x x', a_y y', a_z z') \quad , \qquad (14)$$

where $x', y', z'$ represent a coordinate system with the trap center at the origin. There is an approximate correlation among the coefficients $a_x$, $a_y$, and $a_z$, namely $a_x = -\frac{2}{3}a_z$ and $a_y = -\frac{1}{3}a_z$. The coefficients increase with the bias field, e.g. $a_x$ can be fitted as $9.5 B_{bias}^{1.25}$ G/μm (with $B_{bias}$ evaluated in gauss). However, to estimate the validity of the semiclassical approximation, we make the spherically symmetric approximation $\mu_B \mathbf{S} \cdot \mathbf{B}_{tot} = \alpha \cdot r$, where $\alpha = \mu_B a_x |\mathbf{S}|$, and obtain an analytical solution of Eq. (13). We assume that the spin follows the local direction of the magnetic field, an adiabatic, semiclassical assumption whose validity is examined below. (Here and below **r** is the radius vector for the primed coordinates, i.e. **r** = $(x', y', z')$.) In spherical coordinates, Eq. (13) is rewritten

$$E\psi + \frac{\hbar^2}{2m}\left[\frac{1}{r^2}\frac{\partial}{\partial r}\left(r^2 \frac{\partial \psi}{\partial r}\right) - \frac{\hat{\ell}^2}{r^2}\psi\right] - \alpha r \psi = 0 \quad , \tag{15}$$

where $\hat{\ell}^2$ is the squared angular momentum operator. The solution of (15) is

$$\psi = R(r) Y_{lm}(\theta, \varphi) \quad , \tag{16}$$

where $Y_{lm}(\theta, \varphi)$ are the spherical harmonics. As $\hat{\ell}^2 Y_{lm} = l(l+1) Y_{lm}$ (for $l = 1, 2, \ldots$), the equation for $R(r)$ is

$$\frac{1}{r^2}\frac{\partial}{\partial r}\left(r^2 \frac{\partial R}{\partial r}\right) - \frac{l(l+1)}{r^2} R + \frac{2m}{\hbar^2}[E - \alpha r] R = 0 \quad . \tag{17}$$

Obtaining $R(r)$ as $R(r) = \chi(r)/r$, we rewrite Eq. (17) as

$$\frac{d^2 \chi}{dr^2} + \left[\frac{2m}{\hbar^2}(E - \alpha r) - \frac{l(l+1)}{r^2}\right]\chi = 0 \quad . \tag{18}$$

The $l = 0$ solution of Eq. (17) is

$$R(r) = c_1 \frac{Ai(\eta)}{r} + c_2 \frac{Bi(\eta)}{r} \quad , \tag{19}$$

where $Ai$, $Bi$ are the Airy functions and $\eta = \left(\frac{2m}{\hbar^2 \alpha^2}\right)^{1/3}(\alpha r - E)$. From the requirement that $R(r)$ be bounded for $r \to 0$ and $r \to \infty$ we obtain $c_2 = 0$, and the bound-state energies $E_n$ are given by the zeros of $Ai$:

$$Ai\left[-E_n\left(\frac{2m}{\hbar^2 \alpha^2}\right)^{1/3}\right] = 0. \tag{20}$$

These zeros are tabulated in [28] or can be found numerically; the first few are

$$E_n\left(\frac{2m}{\hbar^2 \alpha^2}\right)^{1/3} = 2.338; 4.088; 5.521; 6.787; 7.944; 9.023; 10.04, \ldots \quad . \tag{21}$$

From (21) we obtain $\Delta E = E_n - E_{n-1} \approx \left(\frac{2m}{\hbar^2 \alpha^2}\right)^{-1/3}$ and the criterion for applying a semiclassical thermodynamic treatment is

$$\Delta E / k_B T = \frac{1}{k_B T} \left(\frac{\hbar^2 \alpha^2}{2m}\right)^{1/3} \ll 1 \quad , \tag{22}$$

where $k_B = 1.38 \times 10^{-23}$ J/K is the Boltzmann constant and $T$ is the temperature of the atoms. That is, the average kinetic energy of the atoms must be large compared to the level spacing, for a thermodynamic treatment.

To estimate the ratio $\Delta E/k_B T$, we set the atomic magnetic moment equal to the Bohr magneton $\mu_B = 9.274 \times 10^{-28}$ J/G and the mass to the $^{87}$Rb mass $m = 1.443 \times 10^{-25}$ kg. The ratio $\Delta E/k_B T$ decreases with temperature. At a bias field of 6 G at $T = 1$ µK, the ratio equals 45, while at the highest temperature determined by $k_B T = \mu_B B_{bias}$, the ratio is about 0.1. (As Sect. 2 shows, the trap depth and the bias field can be taken equal.) So a semiclassical treatment is valid at higher temperatures. In this treatment, atoms leave the trap in two cases: first, if their kinetic energy is larger than the trap depth; second, as a result of spin flips due to non-adiabaticity.

For the first case, we estimate the rate of escape from the trap via the Boltzmann factor, neglecting the details of the trapping potential and treating it as a square well. The Maxwell-Boltzmann distribution for $N_0$ atoms in the trap, with vanishing potential, is

$$dN = \frac{4N_0}{\sqrt{\pi}\bar{u}^3} \exp\left(-\frac{u^2}{\bar{u}^2}\right) u^2 du \quad , \tag{23}$$

where $dN$ is the density of atoms with speed $u$ and $\bar{u} = \sqrt{2k_B T/m}$ is the most probable velocity. We assume that this distribution always applies. If $V_0$ is the trap depth, an atom with kinetic energy $mu^2/2 > V_0$ can leave the trap. The time required for an atom with speed $u$ to leave the trap is $w/u$, where $w$ is the radius of the trap. Thus the total escape rate $\Gamma$ from the trap is

$$\Gamma = \frac{4N_0}{w\sqrt{\pi}\bar{u}^3} \int_{u_{min}}^{\infty} u \exp\left(-\frac{u^2}{\bar{u}^2}\right) u^2 du = \frac{2\bar{u} N_0}{w\sqrt{\pi}} \left(1 + \frac{u_{min}^2}{\bar{u}^2}\right) \exp\left(-\frac{u_{min}^2}{\bar{u}^2}\right) \quad , \tag{24}$$

where $u_{min} = \sqrt{2V_0/m}$. The average escape rate is given by Eq. (24) divided by $N_0$ (so we can say the average escape rate is the average speed divided by $w$) and the average lifetime $\tau_{ad}$ is the inverse of the average escape rate. (Note $u_{min}/\bar{u} = \sqrt{V_0/k_B T}$.) We also consider dependence of the lifetime on temperature. Eq. (24) implies

$$\tau_{ad} = \frac{e^{u_{min}^2/\bar{u}^2} w\sqrt{\pi}}{2\bar{u}\left(1 + u_{min}^2/\bar{u}^2\right)} \quad ; \tag{25}$$

if we fix the trap depth $V_0$ while allowing $T$ to vary in $\bar{u} = \sqrt{2k_B T/m}$ and consider low temperatures, the behavior of $\tau_{ad}$ is

$$\tau_{ad} \sim \sqrt{T} \exp(mu_{min}^2 / 2k_B T) \quad . \tag{26}$$

Another process inducing atom loss is the nonadiabatic (Majorana) spin flips of atoms passing too close to the zero field in the trap center. Petrich et al. [32] assumed that some of atoms cannot adjust their direction to remain parallel to the local magnetic field as they move through the trap, and they considered an atom moving with speed $v$ past the center of the trap with impact parameter $b$. They applied the criterion that the atom's Larmor frequency $b\alpha/\hbar$ must be larger than $v/b$, the maximum rate of change in the direction of the magnetic field acting on the atom; thus $b^2 \geq \hbar v/\alpha$. The radius $r_{max}$ of trapped atoms is given by equipartition, namely $\alpha r_{max} = mv^2/2 = k_B T$, hence the density of the atoms is $n \approx (\alpha/k_B T)^3 N$. The number of spin-flipped atoms that escape from the trap is the flux $nv$ times the cross section $\approx b^2$ of the non-adiabatic volume. Hence the spin flip rate per particle is $1/\tau_{sf} \approx nvb^2/N$ and we have $\tau_{sf} \approx \frac{m}{\hbar}(k_B T/\alpha)^2$. This lifetime increases with $T$ due to a decreased probability for an atom to come within $b$ of the trap center.

Fig. 5 shows the dependence of the lifetimes $\tau_{ad}$ and $\tau_{sf}$ on temperature for two cases: a bias field 0.6 G in Fig. 5(a) and 0.0033 G in Fig. 5(b). Application of the latter bias field creates a trap with the minimal possible gradient of the total magnetic field, 30 G/cm. Temperature is normalized by the temperature $T_0$ at which $\Delta E/k_B T_0 = 1$, namely $T_0 = 3.6$ μK and $T_0 = 47$ nK for bias fields 0.6 G and 0.0033 G, respectively.

Both the optimal temperature and the effective lifetime (trapping time) $\tau$ can be determined by the points where the graphs cross in Figs. 5(a) and 5(b). At $3.4\, T_0 = 12$ μK, corresponding to a bias field of 0.6 G, the effective lifetime is $\tau = 0.05$ ms; at $1.7\, T_0 = 80$ nK, corresponding to a bias field of 0.0033 G, the effective lifetime is $\tau = 1.0$ ms. The lifetimes estimated using a quantum mechanical treatment at $T = 0$ (see the next section) are shown as well, via black arrows. The quantum mechanical treatment yields lifetimes very similar to those predicted by our semiclassical treatment at the optimal temperature. For all the mechanisms considered, the estimated lifetime increases with decrease in gradient of the total magnetic field inside the trap. Atom loss caused by the nonadiabatic spin flip dominates at low temperatures as shown in Fig. 5.

If the temperature $T$ of the atoms satisfies $T < \frac{1}{k_B}\left(\frac{\hbar^2 \alpha^2}{2m}\right)^{1/3}$ (which implies $T < 3.5$ μK for the case of $B_{bias} = 0.6$ G), practically all the atoms will populate the zero energy level. The quantum mechanical treatment for this case is developed in the next section.

**4. Trap stability: quantum mechanical treatment**

Let us now reconsider the applicability of the adiabatic approximation, returning to Eq. (13), the Schrödinger equation for an atom with magnetic moment $\mu_B \mathbf{S}$ in the magnetic trap. We write the potential energy as $V = \mu_B \mathbf{S} \cdot \mathbf{B}$. (For simplicity, here we drop the subscript in $\mathbf{B}_{tot}$.) According to Eq. (14) and the sentence following, $\mathbf{B}$ is proportional to $2\mathbf{x}'/3 + \mathbf{y}'/3 -$

**z′**, but subsequently Sect. 3 takes **B** proportional to $r$, the norm of $(x', y', z')$. Here we take **B** proportional to **x′** + **y′** − 2**z′** which, though not exact, is a better than the spherically symmetric approximation, and preserves Maxwell's equation $\nabla \cdot \mathbf{B} = 0$. We thus admit the possibility of spin flips. We also, for convenience, drop the primes on the coordinates, letting $(x, y, z)$ represent displacement from the minimum of the trap. For simplicity, we specialize to the case of spin-½; the behavior of higher spins should be qualitatively similar. Then the potential energy $V = \mu_B \boldsymbol{\sigma} \cdot \mathbf{B}/2$ for the vortex trap is proportional to

$$r\begin{pmatrix} -2\cos\theta & e^{-i\varphi}\sin\theta \\ e^{i\varphi}\sin\theta & 2\cos\theta \end{pmatrix} , \tag{27}$$

where $\boldsymbol{\sigma} = (\sigma_x, \sigma_y, \sigma_z)$ are the Pauli matrices. The eigenvalues are proportional to $\pm r\sqrt{1+3\cos^2\theta}$ and the normalized eigenvectors are

$$|+\rangle = \frac{1}{\sqrt{2}\sqrt{1+3\cos^2\theta+2\cos\theta\sqrt{1+3\cos^2\theta}}}\begin{pmatrix} \sin\theta \\ e^{i\phi}(2\cos\theta+\sqrt{1+3\cos^2\theta}) \end{pmatrix}, \quad |-\rangle = i\sigma_y|+\rangle . \tag{28}$$

Let $\mathbf{B} = B_0\, r\boldsymbol{\rho}$, where $\boldsymbol{\rho} = (x/r, y/r, -2z/r) = (\sin\theta\cos\varphi, \sin\theta\sin\varphi, -2\cos\theta)$. Refs. [36-38] develop a general method for deriving an effective adiabatic Hamiltonian $H^{eff}$ for a Hamiltonian $H$ by sandwiching it between projectors (projection operators) onto the eigenstates of the "fast" part of $H$. In our case, the "fast" part of $H$ is $V$, and the eigenstates of $V$ are the states $|+\rangle$ and $|-\rangle$ above. The corresponding projectors then are $\Pi_+ = |+\rangle\langle+|$ and $\Pi_- = |-\rangle\langle-|$. Thus, an effective adiabatic Hamiltonian $H_+^{eff}$ for the state $|+\rangle$ can be obtained from the exact Hamiltonian $H$ by sandwiching $H$ between the projector $\Pi_+ = |+\rangle\langle+|$. Since

$$H = p^2/2m + \mu_B \boldsymbol{\sigma} \cdot \mathbf{B}/2 = p^2/2m + \mu_B B_0 r \boldsymbol{\sigma} \cdot \boldsymbol{\rho}/2 , \tag{29}$$

the effective Hamiltonian for the state $|+\rangle$ is $H_+^{eff} = \Pi_+\left(p^2/2m\right)\Pi_+ + \Pi_+(\mu_B B_0 r/2)\boldsymbol{\sigma}\cdot\boldsymbol{\rho}\Pi_+$; **p** is the momentum of an atom. We can write $\Pi_\pm = 1/2 \pm \boldsymbol{\sigma}\cdot\boldsymbol{\rho}/2\sqrt{1+3\cos^2\theta}$ since the matrix $\boldsymbol{\sigma}\cdot\boldsymbol{\rho}$ applied to $|\pm\rangle$ yields the eigenvalues $\pm\sqrt{1+3\cos^2\theta}$. Let us now express **p** as a sum of two parts: $\mathbf{p} = \mathbf{p} - \mathbf{A} + \mathbf{A}$, where $\mathbf{p} - \mathbf{A}$ is purely diagonal and **A** is purely off-diagonal. Namely, $\mathbf{p} - \mathbf{A}$ satisfies $[\mathbf{p} - \mathbf{A}, \Pi_n] = 0$ and **A** satisfies $\Pi_n \mathbf{A} \Pi_n = 0$, for $n = \pm$. We can write $p^2$ as the sum

$$p^2 = (\mathbf{p} - \mathbf{A} + \mathbf{A})^2 = (\mathbf{p} - \mathbf{A})^2 + \mathbf{A}^2 + (\mathbf{p} - \mathbf{A})\mathbf{A} + \mathbf{A}(\mathbf{p} - \mathbf{A}) , \tag{30}$$

and the $p^2$ term in the effective Hamiltonian for the state $|+\rangle$ is then

$$\Pi_+ p^2 \Pi_+ = (\mathbf{p} - \mathbf{A})^2 \Pi_+ + \Pi_+ \mathbf{A}^2 \Pi_+ , \tag{31}$$

with $H' = (\mathbf{p} - \mathbf{A})\mathbf{A} + \mathbf{A}(\mathbf{p} - \mathbf{A})$ dropping out since it is purely off-diagonal.

We recognize **A** as an induced vector potential [39]. But note that the adiabatic approximation induces also a *scalar* potential $\Pi_+ \mathbf{A}^2 \Pi_+ / 2m$ in the state $|+\rangle$.

Now Eq. (8) in Ref. [36] implies that $\mathbf{A} = [\Pi_+, [\Pi_+, \mathbf{p}]]/2 + [\Pi_-, [\Pi_-, \mathbf{p}]]/2$, but these two terms are equal; so we can compute **A** as $\mathbf{A} = [\Pi_+, [\Pi_+, \mathbf{p}]]$ as follows:

$$\mathbf{A} = [\Pi_+, [\Pi_+, \mathbf{p}]] = \frac{i\hbar}{4\sqrt{1+3\cos^2\theta}} \left[ \boldsymbol{\sigma}\cdot\boldsymbol{\rho}, \nabla \frac{\boldsymbol{\sigma}\cdot\boldsymbol{\rho}}{\sqrt{1+3\cos^2\theta}} \right] . \quad (32)$$

The gradient in Eq. (32) reduces to two terms, but one of the terms is proportional to $\boldsymbol{\sigma}\cdot\boldsymbol{\rho}$ and thus doesn't contribute to the commutator. Writing $r\boldsymbol{\rho} = (x, y, -2z)$, we get

$$\mathbf{A} = \frac{i\hbar}{4r(1+3\cos^2\theta)} [\boldsymbol{\sigma}\cdot\boldsymbol{\rho}, (\sigma_x, \sigma_y, -2\sigma_z)] , \quad (33)$$

and the components are

$$A_x = \frac{\hbar(y\sigma_z + 2z\sigma_y)}{2r^2(1+3\cos^2\theta)} , \quad A_y = -\frac{\hbar(x\sigma_z + 2z\sigma_x)}{2r^2(1+3\cos^2\theta)} , \quad A_z = \frac{\hbar(2y\sigma_x - 2x\sigma_y)}{2r^2(1+3\cos^2\theta)} . \quad (34)$$

Therefore $\mathbf{A}^2/2m = \frac{\hbar^2(5x^2 + 5y^2 + 8z^2)}{8mr^4(1+3\cos^2\theta)^2} = \frac{\hbar^2(5+3\cos^2\theta)}{8mr^2(1+3\cos^2\theta)^2}$ is the induced scalar potential felt by the atom in the $|+\rangle$ state (and also in the $|-\rangle$ state). (Note, $\mathbf{A}^2$ is diagonal in the $|+\rangle, |-\rangle$ space and contains an implicit $2 \times 2$ identity matrix.) For $\cos^2\theta$ between 0 and 1, the factor $\frac{5+3\cos^2\theta}{(1+3\cos^2\theta)^2}$ is always positive and drops from 5 (at $\cos^2\theta = 0$) to 1/2 (at $\cos^2\theta = 1$), and the induced potential is always *repulsive*. It does not have radial symmetry but it has rotational symmetry around the *z*-axis (reflecting the radial symmetry of **B** in our approximation).

The induced vector potential **A** yields an effective magnetic field [39]. We can calculate it most easily using Eq. (14) of Ref. [36]:

$$F^{eff}{}_{jk} = \frac{i}{\hbar} \sum_n \Pi_n [A_j, A_k] \Pi_n , \quad (35)$$

i.e. $F^{eff}{}_{jk}$ is $i/\hbar$ times the diagonal part of $[A_j, A_k]$. In our case the commutator is itself diagonal and we have the following effective (induced) $\mathbf{B}^{eff}$:

$$\mathbf{B}^{eff}(\mathbf{r}) = \frac{\hbar\,\boldsymbol{\sigma}\cdot\boldsymbol{\rho}}{r^3(1+3\cos^2\theta)^2} \mathbf{r} . \quad (36)$$

For example, the calculation of $B^{eff}_x$ is

$$B_x^{eff} = F_{yz} = \frac{i}{\hbar}[A_y, A_z] = \frac{i\hbar}{4r^4(1+3\cos^2\theta)^2}[-x\sigma_z - 2z\sigma_x, y\sigma_x - x\sigma_y]$$

$$= \frac{i\hbar}{4r^4(1+3\cos^2\theta)^2}(-4i)(xr\,\boldsymbol{\sigma}\cdot\boldsymbol{\rho}) = \frac{\hbar x\,\boldsymbol{\sigma}\cdot\boldsymbol{\rho}}{r^3(1+3\cos^2\theta)^2} \quad,$$

(37)

and the calculations for $B_y^{eff}$ and $B_z^{eff}$ are similar. In the state $|\pm\rangle$, the matrix $\boldsymbol{\sigma}\cdot\boldsymbol{\rho}$ takes the eigenvalue $\pm\sqrt{1+3\cos^2\theta}$, hence for the $|+\rangle$ state we can write

$$\mathbf{B}_+^{eff}(\mathbf{r}) = \frac{\hbar}{r^3(1+3\cos^2\theta)^{3/2}}\mathbf{r} \quad, \tag{38}$$

namely the induced magnetic field is radial, like a magnetic monopole field, but with additional dependence on the angle $\theta$. The divergence of $\mathbf{B}_+^{eff}$ vanishes.

It is not practical to compute and solve the exact effective Hamiltonian, including the effective vector potential. But we can estimate two kinds of ground-state energies. First, the effective potential we have obtained is

$$V^{eff} = \frac{\hbar^2}{8mr^2}\frac{5+3\cos^2\theta}{(1+3\cos^2\theta)^2} + \frac{\mu_B B_0}{2}r\sqrt{1+3\cos^2\theta} \quad, \tag{39}$$

which we can minimize to obtain $\bar{r}$, the bottom of the potential, and $\bar{\omega}$, the angular frequency of small oscillations near this minimum. Second, we can calculate the angular frequency $\bar{\omega}^{eff}$ of cyclotron motion in the effective field $\mathbf{B}^{eff}$, at a distance $\bar{r}$ from origin. We can then calculate the corresponding ground-state energies $\hbar\bar{\omega}/2$ and $\hbar\bar{\omega}^{eff}$. For these calculations we can drop such factors as $1+3\cos^2\theta$, since all we expect is order-of-magnitude estimates for $\bar{r}$, $\bar{\omega}$, $\bar{\omega}^{eff}$ etc. We have

$$0 = \left.\frac{\partial V^{eff}}{\partial r}\right|_{\bar{r}} = -\frac{\hbar^2}{4m\bar{r}^3} + \frac{\mu_B B_0}{2} \quad, \tag{40}$$

so $\bar{r} = \left(\hbar^2/2m\mu_B B_0\right)^{1/3}$. The second derivative there is $\left.\partial^2 V^{eff}/\partial r^2\right|_{\bar{r}} = 3\hbar^2/4m\bar{r}^4$; equating this second derivative with $m\bar{\omega}^2$, we have $\bar{\omega} = \sqrt{3}(\hbar/2m\bar{r}^2) = \sqrt{3}(\mu_B^2 B_0^2/2m\hbar)^{1/3}$ and ground-state energy $E_0 = \hbar\bar{\omega}/2 = \sqrt{3}(\mu_B^2 B_0^2 \hbar^2/4m)^{1/3}$. For the cyclotron frequency, we have $\omega^{eff} = B^{eff}/m \approx \hbar/m\bar{r}^2$ and energy $\hbar^2/m\bar{r}^2$.

So far, we have assumed adiabaticity and derived the adiabatic Hamiltonian, but we have not determined the range of validity of the adiabatic assumption. We now use Fermi's Golden Rule to find the rate $\Gamma$ of decay from the ground (trapped) state $|g\rangle$ (spin state $|+\rangle$) to the free state $|f\rangle$ (spin state $|-\rangle$) under the influence of the non-diagonal part $H'$ of $H$:

$$\Gamma = \frac{2\pi}{\hbar}|\langle f|H'|g\rangle|^2 \rho(E) \quad, \tag{41}$$

with $H' = (\mathbf{p} - \mathbf{A})\mathbf{A} + \mathbf{A}(\mathbf{p} - \mathbf{A})$ as defined above and $\rho(E)$ as the density of states at energy $E$. There are three steps in the calculation of $\Gamma$: deriving the states $|g\rangle$ and $|f\rangle$, calculating the matrix element in Eq. (41), and computing the density of states $\rho(E)$.

From the start, we can replace $H'$ with $H$ since the matrix element anyway yields only the off-diagonal part. To simplify, we will also assume spherical symmetry. The effective Hamiltonian $H_+^{eff}$ for the trapped wave function $\psi_g$ is

$$H_+^{eff} = -\frac{\hbar^2}{2m}\nabla^2 + \frac{\hbar^2}{8mr^2} + \frac{\mu_B B_0}{2} r \quad . \tag{42}$$

The Schrödinger equation is $E_0 \psi_g = H_+^{eff} \psi_g$, and for $\psi_g = \psi_g(r)$ it is

$$E_0 \psi_g = -\frac{\hbar^2}{2mr}\frac{\partial^2}{\partial r^2}(r\psi_g) + \left(\frac{\hbar^2}{8mr^2} + \frac{\mu_B B_0}{2} r\right) \psi_g \quad . \tag{43}$$

We can define $u_g(r) = r\psi_g(r)$ to get an effective one-dimensional Schrödinger equation:

$$E_0 u_g(r) = -\frac{\hbar^2}{2m}\frac{\partial^2}{\partial r^2} u_g(r) + \left(\frac{\hbar^2}{8mr^2} + \frac{\mu_B B_0}{2} r\right) u_g(r) \quad . \tag{44}$$

We have already computed the minimum of this potential: it occurs at $r = \bar{r}$, and the potential corresponds to small oscillations of angular frequency $\bar{\omega}$ around this minimum. Thus we can approximate $u_g(r)$ by the normalized wave function of a one-dimensional harmonic oscillator of angular frequency $\bar{\omega}$ and centered at $r = \bar{r}$:

$$u_g(r) = \frac{1}{\sqrt{4\pi}} (m\bar{\omega}/\pi\hbar)^{1/4} e^{-m\bar{\omega}(r-\bar{r})^2/2\hbar} \quad , \tag{45}$$

where the additional normalization factor $1/\sqrt{4\pi}$ is due to the definition $u_g(r) = r\psi_g(r)$ which implies $1 = \int_0^\infty 4\pi r^2 \psi_g^2 dr = \int_0^\infty 4\pi u_g^2 dr$. (The normalization is only approximate, since integration yields $\frac{\sqrt{\pi\hbar}}{2\sqrt{m\bar{\omega}}}\left[1 + \text{Erf}\left(\bar{r}\sqrt{\frac{m\bar{\omega}}{\hbar}}\right)\right] \approx \frac{\sqrt{\pi\hbar}}{\sqrt{m\bar{\omega}}}$, where $\bar{r}\sqrt{\frac{m\bar{\omega}}{\hbar}} \approx 0.72$.)

In the Hamiltonian for the free wave functions $u_f(r)$ and $r\psi_f(r)$, the only difference is that the linear part of the potential changes sign, i.e. $H_+^{eff}$ and $H_-^{eff}$ differ only in the relative sign of the term $\mu_B B_0 r/2$. We can use the WKB approximation [40] to estimate $u_f(r)$; namely, $u_f(r)$ is proportional to $[k_f(r)]^{-1/2}$ times an $r$-dependent phase $i\int^r k_f(r')dr'$, where

$$k_f(r) = \sqrt{2m[E_0 - V_f(r)]}/\hbar = \frac{1}{\hbar}\sqrt{2m\left[E_0 - \left(\frac{\hbar^2}{8mr^2} - \frac{\mu_B B_0}{2} r\right)\right]} \quad . \tag{46}$$

Let us consider the normalization of $u_f(r)$. Suppose that we normalize within a symmetric sphere of radius $R$, where $R$ is arbitrarily large, i.e. we must take the limit as $R$ becomes infinite since the sphere is a fiction. For the normalization, the $r$-dependent phase does not matter and we have $|u_f(r)|^2$ proportional $1/k_f(r)$. For large $r$, a good approximation to $|u_f(r)|^2$ is $|u_f(r)|^2 \approx r^{-1/2}$. In this range, $u_f(r)$ approaches an Airy function: $u_f(r) \approx r^{-1/4} \exp(\pm \frac{2}{3} i r^{3/2})$. The integral of $|u_f(r)|^2$ is thus dominated by $\sqrt{R}$ and, in the limit of infinite $R$, only this term remains (up to constant factors). Thus $u_f(r)$ contains a factor $R^{-1/4}$. More precisely, we write

$$u_f(r) = \frac{[\text{phase}]}{\sqrt{8\pi}\left(2E_0/\mu_B B_0 - \hbar^2/4m\mu_B B_0 r^2 + r\right)^{1/4} R^{1/4}} \quad , \tag{47}$$

and it is easy to check that $4\pi \int^R |u_f|^2 \, dr$ approaches 1 in the limit $R \to \infty$.

To apply Fermi's rule, we must calculate the density of states $\rho(E)$ at an energy $E$. We do so as follows. Assuming still that all the states are isotropic, we can estimate their number as $\rho(E) = \frac{\partial}{\partial E} \int^R k_f(r) dr / 2\pi$, and then

$$\rho(E) = \frac{1}{2\pi\hbar} \frac{\partial}{\partial E} \int^R dr \sqrt{2m\left[E - \left(\frac{\hbar^2}{8mr^2} - \frac{\mu_B B_0}{2} r\right)\right]} \approx \frac{1}{h}\sqrt{\frac{2mR}{\mu_B B_0}} \quad . \tag{48}$$

This $\sqrt{R}$ factor will cancel the $1/\sqrt{R}$ in $|\langle f|H'|g\rangle|^2$ arising from the norm $R^{-1/4}$ of $u_f(r)$ or $\psi_f(r)$ squared. Now since $|\langle f|H'|g\rangle|^2 = |\langle f|H|g\rangle|^2$ and $H$ is $(-\hbar^2 \nabla^2/2m)$, we can replace $|\langle f|H'|g\rangle|^2$ with $|\langle f|H|g\rangle|^2$, where

$$\begin{aligned}\langle f|H|g\rangle &= \int dr (4\pi r^2) \psi_f(r)[-\hbar^2 \nabla^2/2m]\psi_g(r) \\ &= \int dr (4\pi r^2) \psi_f \left[-\frac{\hbar^2}{2mr}\frac{\partial^2}{\partial r^2}(r\psi_g)\right] \\ &= \int dr (4\pi) u_f \left[-\frac{\hbar^2}{2m}\frac{\partial^2}{\partial r^2} u_g\right] = (-4\pi\hbar^2/2m)\int dr\, u_f \left[\frac{\partial^2}{\partial r^2} u_g\right] \quad ,\end{aligned} \tag{49}$$

i.e. we can account for $H'$ just by replacing $u_g(r)$ with $(-4\pi\hbar^2/2m)\partial^2 u_g/\partial r^2$. Thus to compute $|\langle f|H'|g\rangle|$ all we need is $-4\pi\hbar^2/2m$ times the inner product of $u_f(r)$ and $\partial^2 u_g/\partial r^2$, where

$$\partial^2 u_g/\partial r^2 = \frac{1}{\sqrt{4\pi}}(m\bar{\omega}/\pi\hbar)^{1/4}\left[m^2\bar{\omega}^2(r-\bar{r})^2/\hbar^2 - m\bar{\omega}/\hbar\right]e^{-m\bar{\omega}(r-\bar{r})^2/2\hbar} \quad . \tag{50}$$

We could integrate Eq. (49) numerically, but we prefer to first find its dependence on $B_0$. It is straightforward to check that $\bar{r}$ is proportional to $(B_0)^{-1/3}$, $\bar{\omega}$ and $E_0$ are proportional to $(B_0)^{2/3}$, and also that $\rho(E)$ is proportional to $(B_0)^{-1/2}$. Then by changing the variable of integration from $r$ to $z = (B_0)^{1/3} r$, and likewise from $r'$ to $z' = (B_0)^{1/3} r'$ in the phase integral of $u_f(r)$, we find that the $B_0$-dependence of $u_f(r)$ is $(B_0)^{1/12}$ and the $B_0$-dependence of $\partial^2 u_g/\partial r^2$ is $(B_0)^{5/6}$. (Although the normalization of $u_g(r)$ is not exact, as noted above, this $B_0$-dependence of $\partial^2 u_g/\partial r^2$ is exact since it does not depend on the normalization). In addition, the substitution $dr = (B_0)^{-1/3}dz$ in the integral of Eq. (49) contributes another factor of $(B_0)^{-1/3}$, such that the $B_0$-dependence of $\langle f|H|g\rangle$ comes to $(B_0)^{1/12+5/6-1/3} = (B_0)^{7/12}$. (By contrast, the phase integral in $u_f(r)$ is invariant, since the $B_0$-dependence in the upper limit of integration cancels the $B_0$-dependence of the integrand.) Finally, we deduce from Eq. (41) that the $B_0$-dependence of the decay rate $\Gamma$ is $(B_0)^{2\times(7/12)-1/2} = (B_0)^{2/3}$. The decay rate depends on the magnetic field gradient of the vortex raised to the 2/3 power.

For evaluating the integrals, however, the substitutions $z = \alpha r$ and $z' = \alpha r'$, where $\alpha = \sqrt{m\bar{\omega}/\hbar} = 3^{1/4}2^{-1/6}(m\mu_B B_0/\hbar^2)^{1/3}$, are more convenient. After numerical integration, we obtain the decay rate

$$\Gamma = (0.037)(\mu_B B_0)^{2/3}(\hbar m)^{-1/3} \; , \tag{51}$$

which we can write as $\Gamma = (30 \text{ s}^{-1})(B_0)^{2/3}$ if the units of $B_0$ are G/cm. Thus, according to this quantum calculation, to increase the trapping time $\tau$, we should decrease $B_0$. At $B_0 = 2\times10^4$ G/cm (see Sect. 3) the lifetime (trapping time) $\tau$ is about 45 μs ($\approx$ 0.05 ms); at $B_0 = 3\times10^3$ G/cm, it is $\tau = 0.16$ ms. However, we cannot decrease $B_0$ so much that gravity pulls atoms out of the trap, i.e. we require $\mu_B B_0 > mg$, or $B_0 > 2mg/\mu_B = 30$ G/cm. Thus, the maximum trapping time for $^{87}$Rb atoms in our vortex trap is approximately $\tau = (30 \text{ G/cm})^{-2/3}/(30 \text{ s}^{-1}) = 3.5$ ms.

## 5. Conclusion

We have shown theoretically that the magnetic field of a single vortex, pinned by a superconducting nano-disc of radius roughly 100 nm and combined with an external bias field, yields a closed 3D trap for cold atoms. The significant advantage of our trap is that technical noise is eliminated, since there are no transport currents. We studied two mechanisms for decay: decay due to thermal escape and due to spin flips (the Majorana instability). These two semiclassical decay mechanisms of Sect. 3 cross in Fig. 5, yielding $\tau = 0.05$ ms in Fig. 5(a) at the optimal temperature $T = 12$ μK (for $B_0 = 2\times10^4$ G/cm), and $\tau = 1.0$ ms in Fig. 5(b) at the optimal temperature $T = 80$ nK (for $B_0 = 30$ G/cm). We compare these *semiclassical* estimates of $\tau$ with the *quantum* estimates of $\tau$ at $T = 0$ obtained in Sect. 4. According to Eq. (51) of Sect. 4, the quantum and semiclassical estimates of $\tau$ coincide in Fig. 5(a), while in Fig. 5(b) the quantum estimate is 3.5 times the semiclassical estimate.

These results demonstrate the possibility of a nano-trap with a height of several tens or hundreds of nanometers above a superconducting chip surface. As the trap approaches the surface, its bias field $B_{bias}$ and field gradient $B_0$ increase and its lifetime decreases. We therefore consider our traps as practical for $B_{bias} \leq 0.6$ G at a height of at least 150 nm, corresponding to a lifetime $\tau \geq 0.05$ ms at low temperatures (Fig. 5). We note that the Casimir-Polder force (Eqs. (37-38) of Ref. [41]) is much smaller than the forces in our trap at heights above 100 nm.

**Acknowledgements**

The authors would like to thank L. Prigozhin, D. Cohen and R. Folman for valuable discussions, and the authors of Ref. [13] for bringing their work to our attention. We acknowledge the support of the German-Israeli Foundation for Scientific Research (GIF).

**Figures**

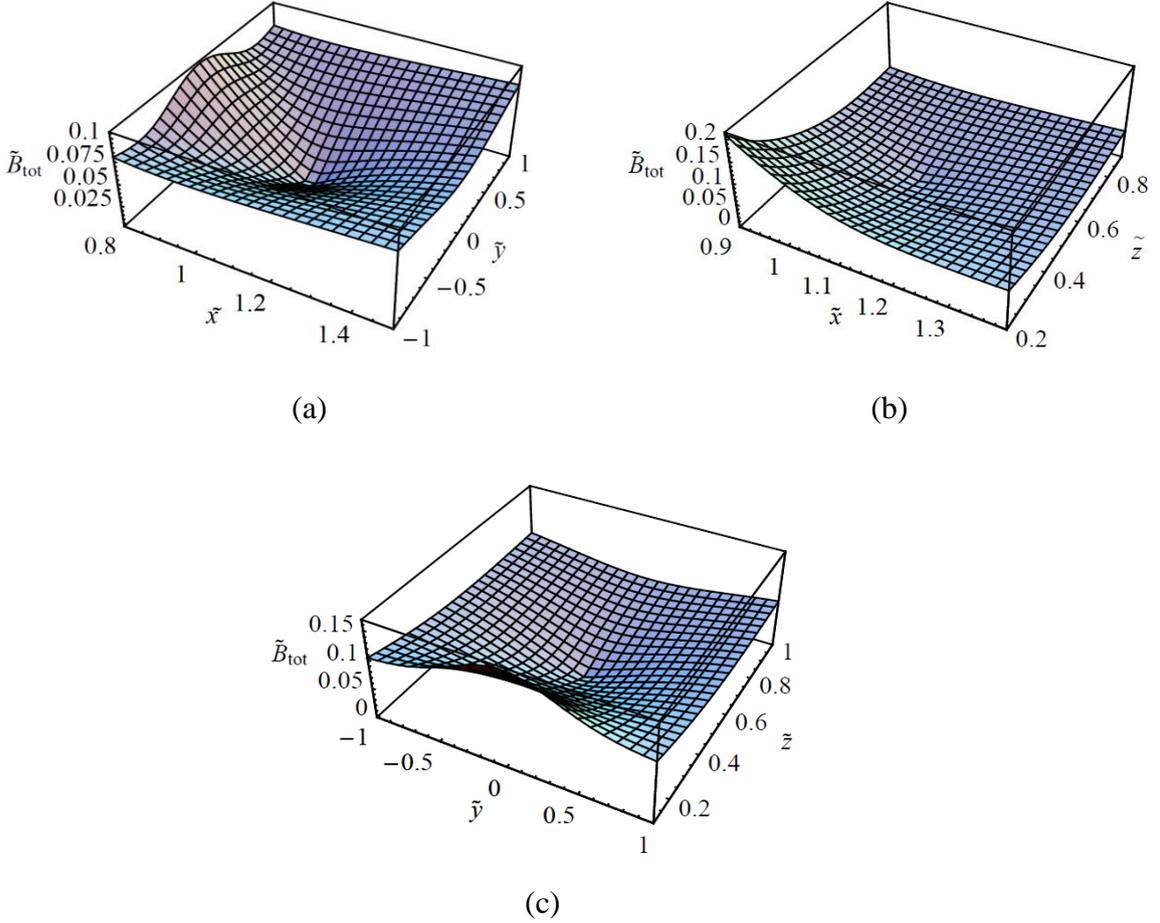

(a)  (b)

(c)

Figure 1. Magnetic trap created by combining a single vortex pinned by a superconducting disc (of radius $R$ and thickness $\delta \ll R$) with a bias field parallel to the disc surface; we define $\tilde{x} = x/R$, $\tilde{y} = y/R$, $\tilde{z} = z/R$: (a) Magnitude of the total magnetic field as a function of $\tilde{x}$ and $\tilde{y}$ at $\tilde{z} = 0.5542$; (b) magnitude of the total magnetic field as a function of $\tilde{x}$ and $\tilde{z}$ at $\tilde{y} = 0$; (c) magnitude of the total magnetic field as a function of $\tilde{y}$ and $\tilde{z}$ at $\tilde{x} = 1.1045$. We plot the magnitude of the magnetic field in units of $\Phi_0 \delta / 2\pi \lambda^2 R$, where $\Phi_0$ is the flux quantum and the bias field, directed opposite to the $x$-axis, equals 0.1. The minimum (trap center) is at (1.1045, 0, 0.5542).

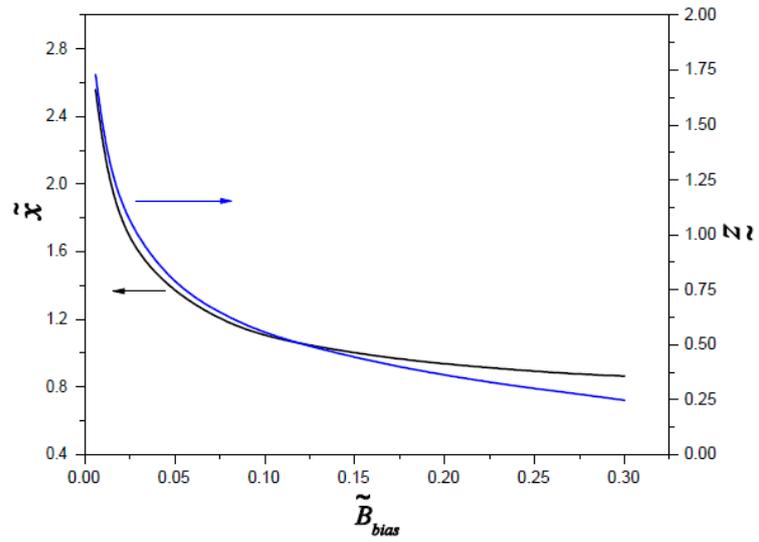

Figure 2 (color online). The $\tilde{x}$- and $\tilde{z}$-coordinates of the trap center as a function of the bias field: $\tilde{x}$-coordinates in black, $\tilde{z}$-coordinates in blue.

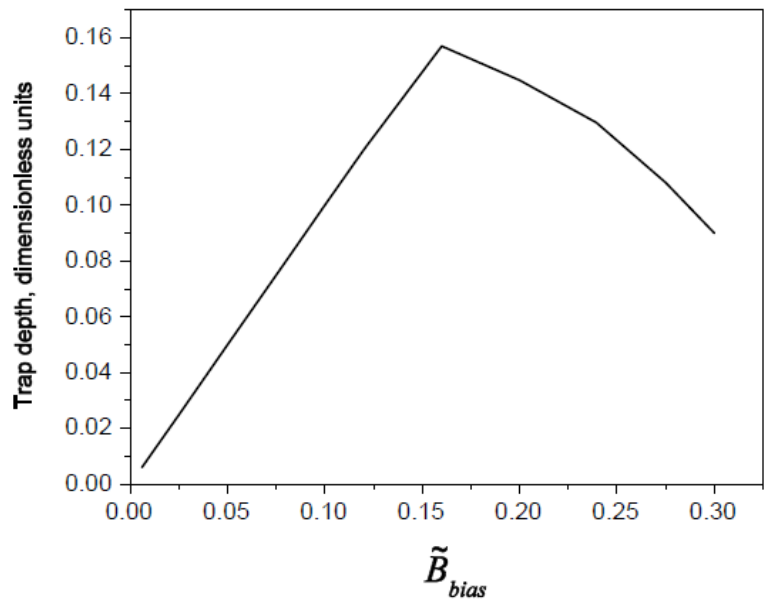

Figure 3. Trap depth as a function of the bias field.

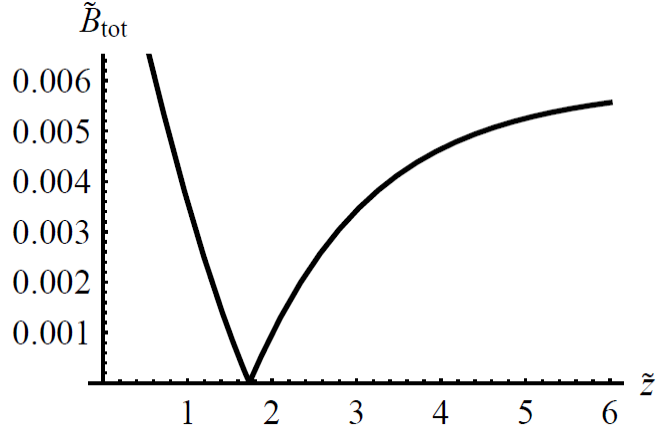

Figure 4. Dependence of the magnitude of the total magnetic field on $\tilde{z}$ at $\tilde{B}_{bias} = 0.006$. The trap center is (2.56, 0, 1.73).

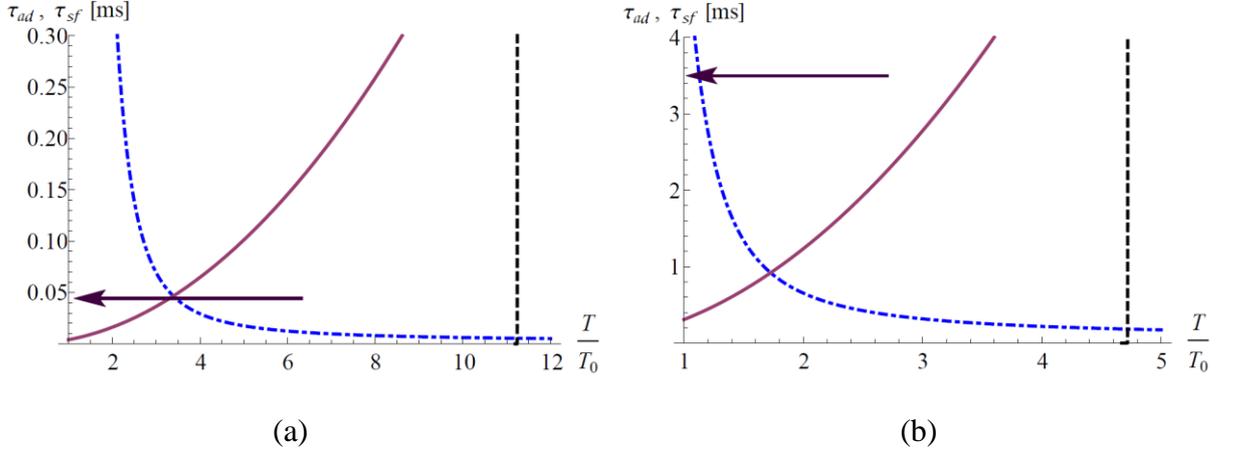

(a)          (b)

Figure 5 (color online). Dependence of the lifetimes $\tau_{ad}$ and $\tau_{sf}$ on temperature for two cases: bias field $B_{bias} = 0.6$ G and gradient $B_0 = 2 \times 10^4$ G/cm in Fig. 5(a); $B_{bias} = 0.0033$ G and $B_0 = 30$ G/cm in Fig. 5(b). Blue dash-dotted curve: thermal escape time $\tau_{ad}$ according to Eq. (25); solid purple curve: nonadiabatic spin flip time, $\tau_{sf}$; black arrow: the quantum estimate of the lifetime at $T = 0$ according to Eq. (51); black dashed line: the trap depth. Temperature is normalized by the temperature $T_0$ at which $\Delta E / k_B T_0 = 1$; see Eq. (22). The lifetime estimates are for a trap above a disc of radius $R = \lambda = 100$ nm and thickness $\delta = \lambda/3$; we have $T_0 = 3.6$ μK in (a) and $T_0 = 47$ nK in (b).